\documentclass[epj,nopacs]{svjour}
\usepackage{latexsym}
\usepackage{graphics}
\usepackage{graphicx}
\usepackage{bm}
\usepackage{epsfig}
\usepackage{amsmath,amsfonts}
\usepackage{url}
\usepackage{epstopdf}
\usepackage{color}
\newcommand{\be}{\begin{equation}}
\newcommand{\ee}{\end{equation}}
\newcommand{\bea}{\begin{eqnarray}}
\newcommand{\eea}{\end{eqnarray}}
\newcommand{\bse}{\begin{subequations}}
\newcommand{\ese}{\end{subequations}}
\newcommand{\bi}{\begin{itemize}}
\newcommand{\nn}{\nonumber}

\begin{document}

\title{The phase space analysis of modified gravity (MOG)}
\author{Sara Jamali\thanks{\emph{e-mail:} sara.jamali@stu.um.ac.ir} \and Mahmood Roshan\inst{*}
\thanks{\emph{e-mail:} mroshan@um.ac.ir}}  
%
%
\institute{Department of Physics, Ferdowsi University of Mashhad, P.O. Box 1436, Mashhad, Iran}
\date{Received: date / Revised version: date}

\date{Received: date / Revised version: date}

\abstract{We investigate the cosmological consequences of a scalar-vector-tensor theory of gravity known as MOG. In MOG, in addition to metric tensor, there are two scalar fields $G(x)$ and $\mu(x)$, and one vector field $\phi_{\alpha}(x)$. Using the phase space analysis, we explore the cosmological consequences of a model of MOG and find some new interesting features which are absent in $\Lambda$CDM model. More specifically we study the possibility that if the extra fields of this theory behave like dark energy to explain the cosmic speedup. More interestingly, with or without cosmological constant, strongly phantom crossing happens. Also we find that this theory in its original form ($\Lambda\neq 0$), possesses a true sequence of cosmological epochs. Albeit we show that, surprisingly, there are two radiation dominated epochs $f_5$ and $f_6$, two matter dominated phases $f_3$ and $f_4$, and two late time accelerated eras $f_{12}$ and $f_{7}$. Depending on the initial conditions the universe will realize only three of these six eras. However, the matter dominated phases are dramatically different from the standard matter dominated epoch. In these phases the cosmic scale factor grows as $a(t)\sim t^{0.46}$ and $t^{0.52}$, respectively, which are slower than the standard case, i.e. $a(t)\sim t^{2/3}$. Considering these results we discuss the cosmological viability of MOG.
\PACS{{PACS-key}{discribing text of that key}}}

\maketitle

\section{Introduction}
\label{intro}
We investigate the cosmological consequences of a modified theory of gravity known as MOG in the relevant literature \cite{moffat2}. MOG is a relativistic theory which exploits three kinds of gravitational fields, i.e. tensor, scalar and vector fields. More specifically, in addition to the metric tensor, MOG possesses two scalar fields $G(x)$, $\mu(x)$ and a Proca vector field $\phi_{\alpha}(x)$. The vector field is directly coupled to the matter fields. Therefore, this theory is not a metric theory of gravity and consequently the weak equivalence principle, in principle, can be violated. Naturally, the free parameters of MOG are chosen such that to make the theory consistent with the experimental tests of the equivalence principle. The main motivation for introducing this theory is to solve the dark matter enigma. It is claimed that MOG can explain the flat rotation curve of the spiral galaxies without adding any dark matter halo \cite{Moffat:2013uaa},  \cite{Brownstein:2005zz} . Also this theory explains the matter discrepancy in the galaxy clusters \cite{Brownstein:2005dr}. It is worthy to mention that, it is not the first time that some modifications in the gravitational law can somehow address the above mentioned problems. For an explicit example we refer the reader to Modified Newtonian Dynamics (MOND) \cite{Milgrom:1983ca} and its relativistic generalizations such as Tensor-Scalar-Vector theory (TeVeS) \cite{Bekenstein:2004ne}. It is recently claimed that MOG is more successful than MOND in explaining the flat rotation curves \cite{Moffat:2014pia}. Also the local stability of spiral galaxies in MOG has been investigated in \cite{Roshan:2015gra}. The gravitational Jeans instability for molecular clouds has been studied in \cite{ra}.

Our purpose in this paper is to study the cosmological behavior of a MOG model. It is important mentioning that like $f(R)$ gravity, MOG may refer to a large class of models corresponding to different energy contributions for the scalar and vector fields. In other words by changing the kinetic and potential energy contributions of the fields, one may construct a new model of MOG. In this paper we restrict ourselves to a MOG model presented in \cite{Roshan:2015uta}. 

The astrophysical consequences of this theory, more specifically astrophysical issues relevant to the dark matter problem, have been widely investigated. Since MOG is still considered as an alternative theory to dark matter particles and has not been ruled out yet, it seems necessary to check its cosmological consequences. We know that adding only a single scalar field to a gravitational theory can lead to significant outcomes in the cosmological issues. For example we recall the quintessence model and Brans-Dicke theory. Therefore it is natural to ask that how is the cosmological behavior of MOG considering the variety of fields that have been incorporated. On the other hand there are a few papers considering cosmology of MOG. For example in \cite{Roshan:2015uta} the Noether symmetries of the cosmic pint-like Lagrangian of MOG has been studied and some exact cosmological solutions have been found. Also in \cite{Moffat:2014bfa} the perturbation growth in the context of MOG has been studied. See \cite{Moffat:2015bda} for relevant works.

In order to check the main cosmological features of MOG, and the minimum requirements that it must possess, we use the dynamical system method (or the phase space analysis). This method provide a fast and reliable procedure to numerically solve the field equations. Note that field equations of MOG are drastically complicated than Einstein's general relativity (GR), see equations \eqref{FE1}-\eqref{FE4}. More importantly, this method enables us to project the dynamics into a compact region and explore the most important "`events"' that can be happen. One of the necessary requirements that a cosmological model should satisfy, is the existence of true sequence of cosmological epochs. More specifically, the cosmic evolution should start with a radiation dominated phase. After this phase there should be a proper matter dominated phase which is long enough to allow the structure formation and fast enough to be consistent with the observations of the age of the universe. Finally the universe should enter an accelerated epoch consistent with the relevant observations such as the Supernovae type Ia data. Fortunately, the dynamical system method is an excellent tool for checking this important requirement. This method has been applied to various alternative theories and cosmological models, for example see \cite{Xu:2012jf}.

This paper is organized as follows. In section \ref{sec-1} we briefly introduce MOG and the modified Friedmann equations. In section \ref{sec-2} we introduce the dynamical system variables and the autonomous first order differential equations. Also we find the critical points and explore their stability and physical relevance. In this section we assume that the cosmological constant is zero. In section \ref{sec-3} we bring back the cosmological constant and analyze the system. In section \ref{sec-4} we study the phase space of the system at infinity. Finally, conclusions are drawn in sec \ref{sec-4.1}.

\section{Modified Friedmann equations in MOG}
\label{sec-1}
Let us start with an action for MOG presented in \cite{Roshan:2015uta}
\begin{equation}\label{action}
\begin{split}
S = &\frac{1}{16 \pi}\int\sqrt{-g}\ d^{4}x 
\bigg[\frac{\chi^2}{2} (R-2\Lambda) +\frac{1}{2} g^{\mu \nu}\nabla_{\mu} \chi \nabla_{\nu} \chi \nonumber
\\&+\frac{\chi^2}{4} g^{\mu \nu} \nabla_{\mu} \psi \nabla_{\nu} \psi +\omega_{0}[\frac{1}{4} B_{\mu \nu}B^{\mu \nu} +V_{\phi}] \bigg] +S_{M}
\end{split}
\end{equation}
where $R$ is the Ricci scalar, $\Lambda$ is
a positive constant corresponding to the cosmological constant in the Einstein-Hilbert action. It is noteworthy that although, in this paper, we will denote $\Lambda$ as the cosmological constant, it can be considered as the mass term for the scalar field $\chi$. In other words, it is not exactly the cosmological constant and one may present different interpretations for its appearance in the action. Also $\omega_{0}$ denotes a positive coupling constant, $S_{M}$ is the matter action and $B_{\mu \nu}=\nabla _{\mu} \phi_{\nu}-\nabla_{\nu}\phi_{\mu}$ is an anti-symmetric tensor reminiscent of the Maxwell's tensor in electrodynamics. The new scalar fields $\chi$ and $\psi$ are related to $G$ and $\mu$ introduced in \cite{moffat2} as $\chi^2=2/G$ and $\psi= \ln \mu$, see \cite{Roshan:2015uta} for more details. It should be stressed that the scalar field $G$, in principle, can be negative. This means that $\chi$ can be a pure imaginary function. Albeit the Lagrangian density in the above action remains always real. However, $\chi$ is an auxiliary function for writing the action in a more common and compact form, and our main scalar field is $G$. The potential $V_{\phi}$ is chosen as $V_{\phi}\propto e^{2\psi} \phi_{\beta}\phi^{\beta}$. This means that $\mu$ appears as a time dependent mass for the vector field and plays a central role for addressing the dark matter problem \cite{moffat2}. 

Varying the action with respect to the fields, one can find the relevant field equations
 \begin{eqnarray} \label{mog11}
 G_{\mu\nu}+\Lambda g_{\mu\nu}=\frac{1}{\chi^2}(\nabla_{\mu}\nabla_{\nu}-g_{\mu\nu}\square)\chi^2+\frac{16\pi}{\chi^2} T^{\text{total}}_{\mu\nu}
 \end{eqnarray}
 \begin{eqnarray}\label{mog12}
 \nabla_{\mu}B^{\alpha\mu}+ \frac{\partial V_{\phi}}{\partial \phi_{\alpha}}=\frac{16\pi}{\omega_0} J^{\alpha}
 \end{eqnarray}
 \begin{eqnarray}\label{mog13}
    \square \chi=\chi( R-2\Lambda)+\frac{\chi}{2}g^{\mu\nu}\nabla_{\mu}\psi\nabla_{\nu}\psi
 \end{eqnarray}
 \begin{eqnarray}\label{mog14}
 \square \psi=-\frac{2}{\chi}\nabla_{\gamma}\chi\nabla^{\gamma}\psi+\frac{2\omega_0}{\chi^2}\frac{\partial V_{\phi}}{\partial \psi}
 \end{eqnarray}
 Where $G_{\mu\nu}$ is the Einstein tensor and $J^{\alpha}$ is a "fifth force" matter current defined as
 \begin{eqnarray}\label{current1}
J^{\alpha}=-\frac{1}{\sqrt{-g}}\frac{ \delta S_M}{\delta \phi_{\alpha}}
 \end{eqnarray}
 nonzero $J^{\alpha}$ means that there is a coupling between matter and the vector field $\phi^{\mu}$. This coupling can, in principle, lead to a violation of the Einstein's equivalence principle. In this paper we assume that $\nabla_{\alpha}J^{\alpha}=0$. This is an extra assumption and in principle one may study different versions of MOG in which this conservation equation is violated. Also, the total energy-momentum tensor is defined as
\begin{eqnarray}
T^{\text{total}}_{\mu\nu}=T_{\mu\nu}+T^{\phi}_{\mu\nu}+T^{\chi}_{\mu\nu}+T^{\psi}_{\mu\nu}
\end{eqnarray}
where $T_{\mu\nu}$ is the energy-momentum tensor for the ordinary matter, and
\begin{eqnarray}\label{mog7}
\begin{split}
&T^{\phi}_{\mu\nu}=-\frac{\omega_0}{16\pi}\left(B_{\mu}^{~\alpha}B_{\nu\alpha}-g_{\mu\nu}(\frac{B^{\rho\sigma}}{4}B_{\rho\sigma}+V_{\phi})+2\frac{\partial V_{\phi}}{\partial g^{\mu\nu}}\right)\nonumber\\&
T^{\chi}_{\mu\nu}=-\frac{1}{16\pi}\left(\nabla_{\mu}\chi\nabla_{\nu}\chi-\frac{1}{2} g_{\mu\nu}\nabla_{\alpha}\chi\nabla^{\alpha}\chi\right)\nonumber \\&\
T^{\psi}_{\mu\nu}=-\frac{\chi^2}{32 \pi}\left(\nabla_{\mu}\psi\nabla_{\nu}\psi-\frac{1}{2} g_{\mu\nu}\nabla_{\alpha}\psi\nabla^{\alpha}\psi\right)\nonumber
\end{split}
\end{eqnarray}
In order to study the cosmological consequences of MOG, we assume a flat Friedmann-Robertson-Walker (FRW) metric 
$$ds^2=-dt^2+a(t)^2(dx^2+dy^2+dz^2)$$
where $a(t)$ is the cosmic scale factor. Also we assume that the cosmic fluid can be characterized by an ideal fluid with energy density distribution $\rho$, the pressure $p$ and the velocity four vector $u_{\mu}$. In this case the energy momentum tensor is
$$T_{\mu\nu}=(\rho+p)u_{\mu}u_{\nu}+p g_{\mu\nu}$$
Finally, bearing in mind that $\chi^2=2/G$ and $\psi= \ln \mu$, we find the following Friedmann equations
\bea\label{FE10}
\begin{split}
&\frac{\dot{a}^2}{a^2}=\frac{8\pi G}{3}\rho+\frac{\Lambda}{3}\\&+\left[\frac{\dot{G}}{G}\frac{\dot{a}}{a}-\frac{1}{12}\frac{\dot{\mu}^2}{\mu^2}-\frac{1}{24}\frac{\dot{G}^2}{G^2}-\frac{G\omega_0}{3}\left(\frac{V_{\phi}}{2}+\frac{\partial V_{\phi}}{\partial g^{00}}\right)\right]
 \end{split}
\eea
\bea\label{FE20}
\begin{split}
\frac{\ddot{a}}{a}&=-\frac{4\pi G}{3}(\rho+3 p)+\frac{\Lambda}{3}+\left[\frac{1}{2}\frac{\dot{G}}{G}\frac{\dot{a}}{a}+\frac{1}{6}\frac{\dot{\mu}^2}{\mu^2}\right]\\&+\left[\frac{1}{2}\frac{\ddot{G}}{G}-\frac{11}{12}\frac{\dot{G}^2}{G^2}-\frac{G\omega_0}{6}\left(V_{\phi}-\frac{\partial V_{\phi}}{\partial g^{00}}\right)\right]
\end{split}
\eea
\begin{eqnarray}\label{eq3}
\frac{\partial V_{\phi}}{\partial \phi_0}=\frac{16\pi J^0}{\omega_0}
\end{eqnarray}
\bea\label{FE30}
\begin{split}
\frac{\ddot{G}}{G}&=32\pi G \rho+12\frac{\ddot{a}}{a}+9\frac{\dot{G}}{G}\frac{\dot{a}}{a}-2\frac{\dot{\mu}^2}{\mu}+\frac{\dot{G}^2}{G^2}\\&-4 G\omega_0\left(\frac{V_{\phi}}{2}+\frac{\partial V_{\phi}}{\partial g^{00}}\right)
\end{split}
\eea
\bea\label{FE40}
\frac{\ddot{\mu}}{\mu}=\frac{\dot{\mu}^2}{\mu^2}-3\frac{\dot{\mu}}{\mu}\frac{\dot{a}}{a}+\frac{\dot{G}}{G}\frac{\dot{\mu}}{\mu}-G\omega_0\mu\frac{\partial V_{\phi}}{\partial \mu}
\eea
Where a dot stands for a derivative with respect to time $t$, and $\rho$ includes both matter and radiation contributions, i.e. $\rho = \rho_{m} + \rho_{r}$. It is important mentioning that scalar fields $\mu$ and $G$ have negative contributions to the total energy density. More specifically the kinetic terms $\frac{\dot{\mu}^2}{\mu^2}$ and $\frac{\dot{G}^2}{G^2}$ appear with negative sing in \eqref{FE10}. As we will show, this fact leads to some phantom features in this model. It is obvious that if we change the sign of the kinetic terms in the action, then the cosmological consequences of this model, in principle, will change.

In the following we restrict ourselves to the potential $V_{\phi}=-\frac{1}{2} \mu^2 \phi_{\alpha}\phi^{\alpha}$. This is the original potential of MOG presented in \cite{moffat2}. In this case after some algebraic manipulations, we rewrite equations \eqref{FE10}-\eqref{FE40} as follows
\bea\label{FE1}
\begin{split}
\frac{\dot{a}^2}{a^2}=&\frac{8 \pi G}{3} \rho + \frac{\Lambda}{3} + \frac{\dot{G}\dot{a}}{G a} 
 - \frac{\dot{\mu}^2}{12 \mu^2} - \frac{\dot{G}^2}{24 G^2} \\& + \frac{64}{3}\frac{\pi^2 J^2 G}{\omega_{0} \mu^2}
 \end{split}
\eea
\bea\label{FE2}
\begin{split}
\frac{\ddot{a}}{a}=& -\frac{44\pi G }{15}(\rho  -\frac{3}{11}p) -\frac{\Lambda}{15}-\frac{\dot{G}\dot{a}}{G a}+\frac{\dot{G}^2}{12 G^2}
+\frac{\dot{\mu}^2}{6 \mu^2}\\&-\frac{256}{15}\frac{G \pi^2 J^2}{\omega_{0} \mu^2}
\end{split}
\eea
\bea\label{FE3}
\begin{split}
 \frac{\ddot{G}}{G}=& -\frac{16 \pi G}{5} (\rho - 3p) -\frac{4 \Lambda}{5} +\frac{2 \dot{G}^2}{G^2}-\frac{3 \dot{G} \dot{a}}{G a}
\\&+\frac{256}{5}\frac{G \pi^2 J^2}{\omega_{0} \mu^2}
\end{split}
\eea
\bea\label{FE4}
\frac{\ddot{\mu}}{\mu}&=&\frac{\dot{\mu}^2}{\mu^2}-3\frac{\dot{\mu}\dot{a}}{\mu a}+\frac{\dot{G}\dot{\mu}}{G \mu } -\frac{256 \pi^2 G J^2}{\omega_0 \mu^2}
\eea
Note that using the field equation of the vector field we have replaced $\phi_0$, the only non-zero component of the vector field, with $\frac{16 \pi J(t)}{\omega_0 \mu^2}$, Where $J=J^0$ is the time component of the matter current $J^{\alpha}$. It is clear that these non-linear differential equations, i.e. equations \eqref{FE1}-\eqref{FE4}, are drastically complicated than the standard Friedmann equations. However, as we shall see, despite of this complexity the dynamical system approach provides a fast numerically stable integration of the equations.

As we have already mentioned the equivalence principle can be violated in this theory. Consequently the ordinary energy-momentum tensor $T_{\mu\nu}$ is not conserved \cite{Roshan:2014mqa}. However, fortunately in the isotropic and homogeneous FRW space-time and with the assumption that $\nabla_{\alpha}J^{\alpha}=0$, $T_{\mu\nu}$ is conserved and one may use the standard relations between energy densities and the scale factor, i.e. $\rho_m\propto a^{-3}$ and $\rho_r\propto a^{-4}$, see \cite{Roshan:2014mqa} for more details. In this case one may straightforwardly set the matter current as $J=\kappa \rho_m$. Where $\kappa$ is an another positive coupling constant. It is noteworthy that in a non-homogeneous space-time because of the coupling between matter and the vector field, these relations are not true and one may expect significant departures from $\Lambda$CDM model. We recall that there are several attempts in the literature to find a relationship between the cosmic-speed up and the inhomogeneities in the distribution of matter, for example see  \cite{Alnes:2005rw} and \cite{GarciaBellido:2008yq}. Therefore, regarding the energy exchange between matter and the vector field in a non-homogeneous background, it seems interesting to check the possibility that if MOG can explain the accelerated expansion without invoking the cosmological constant and just by taking into account the matter inhomogeneities. This issue can be a matter of study for future works. Therefore, in what follows we work in an isotropic and homogeneous background. 

Now let us consider MOG as a dark energy model. In order to find the equation of state parameter of dark energy, i.e. $\omega_{\text{DE}}$, we rewrite equations \eqref{FE1} and \eqref{FE2} as
\begin{equation}
3H^2=8\pi G_{\text{N}}\left(\rho_{m} +\rho_{r}+\rho_{\text{DE}}\right)
\end{equation}
\begin{equation}
-2\dot{H}=8\pi G_{\text{N}} \left(\rho_{m}+\frac{4}{3} \rho_{r} +\rho_{\text{DE}}+p_{\text{DE}}\right)
\end{equation}
where $G_{\text{N}}$ is the Newtonian gravitational constant and $H= \dot{a}/a$ is the Hubble parameter and a dot denotes derivative with respect to cosmic time $t$. Where $\rho_{\text{DE}}$ and $p_{\text{DE}}$ are defined as

\bea
\begin{split}
 8 \pi G_{\text{N}} \rho_{\text{DE}}=& 8\pi(G-G_{\text{N}})\rho +\Lambda +3H\frac{\dot{G}}{G} \\&-\frac{\dot{\mu}^2}{4\mu^2}-\frac{\dot{G}^2}{8G^2}+ 64 \frac{\pi^2 \kappa^2}{\omega_{0}}\frac{G \rho^2}{\mu^2}
 \end{split}
\eea
\bea
\begin{split}
8 \pi G_{\text{N}} p_{\text{DE}} =&\frac{\dot{G}H}{G}  \!-\! \frac{\dot{G}^2}{8 G^2} \!-\! \frac{\Lambda}{5} \!-\! \frac{\dot{\mu}}{4 \mu^2} \! -\! \frac{8 \! \pi (\!G\!-\!G_{N})}{3}\! \rho_{r} 
\\& +\frac{16 \pi G}{5}\rho_{m} \!+\!  \frac{1216 \pi ^2 G \kappa ^2 \text{$\rho_{ m}$}^2}{15 \mu ^2 \omega }
 \end{split}
\eea

Now it is possible to express the equation of state parameter of dark energy by noticing that $\omega_{\text{DE}} =\frac{p_{\text{DE}}}{\rho_{\text{DE}}}$. It is also useful to  write the effective equation of state parameter $\omega_{\text{eff}}$, that conveniently is defined to include all components of the energy budget of the cosmos, namely

\be\label{omegaeff}
\omega_{\text{eff}}=\frac{p_{\text{tot}}}{\rho_{\text{tot}}} = -1-\frac{2\dot{H}}{3H^2}
\ee

\section{Phase space analysis of MoG}
\label{sec-2}
In order to apply the phase-space analysis to MOG, we transform the field equations \eqref{FE1}-\eqref{FE4} into autonomous form $\mathbf{x}' = \mathbf{f(x)}$, where $\mathbf{x}$ is the column vector constituted by an appropriate set of new variables and $\mathbf{f(x)}$ is the corresponding column vector of the autonomous differential equations. Also prime denotes derivative with respect to $\ln a$. The fixed points $\mathbf{x_c}$ of the system satisfy $\mathbf{x}' = 0$, and in order to determine the stability of these points, we perturb the system around the fixed points as
$\mathbf{x} = \mathbf{x_c}+\mathbf{\delta}$, Where $\mathbf{\delta}$ is a column vector for the perturbations. Expanding the autonomous equations  up to linear order in  perturbations, we have $\mathbf{\delta}'=\mathbf{M}~\mathbf{\delta}$ where $\mathbf{M}$ is the stability matrix. Finally, type and the stability of each fixed point, can be found using the eigenvalues of the stability matrix \cite{perko}. Now let us define the following dimensionless variables
\begin{equation}
\begin{split}
& y=\frac{8 \pi G}{3 H^2} \rho_m,   \,\,\ r=\frac{8 \pi G}{3 H^2} \rho_r \,\,\ z =\frac{\dot{G}}{G H}\,\,\, m = \frac{\dot{\mu}}{\sqrt{12}\mu H}\\&
x^2 = \frac{\Lambda}{3 H^2}, \qquad Q=\frac{G}{3\omega}\left(\frac{8 \pi \kappa\rho_m }{ H \mu }\right)^2
\end{split}
\label{var}
\end{equation}
Substituting these dynamical variables into equation \eqref{FE1}, we find a constraint equation
 $$(y + r + z + x^2 - m^2 - \frac{z^2}{24} + Q)=1$$
If we assume that $G>0$, then it is clear form the definition of $Q$ that it is a positive parameter. Therefore we rewrite the constraint equation as
\begin{equation}
y+r+x^2+z-m^2-\frac{z^2}{24} \leq 1
\label{non}
\end{equation}
After some algebraic manipulations, the modified Friedmann equations take the following form
\bea \label{yprime}
y^{\prime}= -\frac{1}{5} 12 m^2 y\!+\!\frac{2 r y}{5}\!-\!\frac{6 x^2 y}{5}\!+\!\frac{3 y^2}{5}\!-\!\frac{y z^2}{10}\!+\!\frac{7 y z}{5}\!+\!\frac{3 y}{5}
\eea
\bea \label{rprime}
r^{\prime}\! = \!-\frac{1}{5} 12 m^2 r\!+\!\frac{2 r^2}{5}\!-\!\frac{6 r x^2}{5}\!+\!\frac{3 r y}{5}\!-\!\frac{r z^2}{10}\!+\!\frac{7 r z}{5}\!-\!\frac{2 r}{5}
\eea

\bea \label{xprime}
x^{\prime} \!= \!-\frac{6 m^2 x}{5}+\frac{r x}{5}-\frac{3 x^3}{5}\!+\!\frac{3 x y}{10}\!-\!\frac{x z^2}{20}\!+\!\frac{x z}{5}\!+\!\frac{9 x}{5}
\eea

\bea\label{zprime}
\begin{split}
z^{\prime}\!\! =&\!\!-\frac{6 m^2 z}{5}\!+\!\!\frac{12 m^2}{5}\!+\!\!\frac{r z}{5}\!-\!\!\frac{12 r}{5}\!-\!\!\frac{3 x^2 z}{5}\!-\!\!\frac{24 x^2}{5}\!+\!\!\frac{3 y z}{10} \\& -\frac{18 y}{5}\!-\!\!\frac{z^3}{20}\!+\!\!\frac{13 z^2}{10}\!-\!\!\frac{18 z}{5}\!+\!\!\frac{12}{5}
\end{split}
\eea

\bea \label{mprime}
\begin{split}
m^{\prime}\!\!=\!\!&-\frac{6 m^3}{5}\!-\!\!2 \sqrt{3} m^2\!+\!\!\frac{m r}{5}\!-\!\!\frac{3 m x^2}{5}\!+\!\!\frac{3 m y}{10}\!-\!\!\frac{m z^2}{20}\!+\!\frac{6 m z}{5}\\&- \frac{6 m}{5}\!+\!\!2 \sqrt{3} r\!+\!\!2 \sqrt{3} x^2\!+\!\!2 \sqrt{3} y\!-\!\!\frac{z^2}{4 \sqrt{3}}\!+\!\!2 \sqrt{3} z\!-\!2 \sqrt{3}
\end{split}
\eea
It is important mentioning that although $G$ (or equivalently $\chi$) can be considered as a time dependent gravitational constant, its sign is not necessarily positive. This means that, in principle, \textit{anti-gravity} is possible in MOG. In fact it is well-known that in non-minimally coupled scalar-tensor theories of gravity, the anti-gravity regime can exist, see \cite{Faraoni} and references therein. For more recent works we refer the reader to \cite{anti}. There is also a non-minimally coupled scalar field $\chi$ in MOG. However, in the following we explicitly show that there is no transitions from anti-gravity to gravity in the context of MOG. More specifically, if the evolution starts from a anti-gravity regime, it will remain permanently at that phase. In other words, if $G$ starts with a negative value, then its sign will not change during the cosmic evolution. Therefore, it has to start form a positive value and one can be sure that $y$ and $r$ are also positive quantities during the whole thermal history. To show this fact more precisely, let us rewrite equation \eqref{FE1} as
\bea\label{Forg1}
\begin{split}
  \frac{H\dot{G}}{|G|}=\,&\text{sgn}(G)\left(H^2+ \frac{\dot{\mu}^2}{12 \mu^2}+\frac{\dot{G}^2}{24 G^2}\right)\\&-\left(\frac{8 \pi |G|}{3} \rho+\frac{64}{3}\frac{\pi^2 \kappa^2 |G| \rho_{m}^2 \phi}{\omega_{0} \mu^2}+\text{sgn}(G)\frac{\Lambda}{3}\right)
  \end{split}
\eea
In the early universe we can neglect the $\Lambda$ term. In this case, supposing a negative value for $G$, one finds that $\dot{G}$ is also negative for an expanding universe. However, there may exist a minimum for $G$ and after that it can increase and finally become positive. Note that there is a positive $\Lambda$ term in the right hand side of \eqref{Forg1}. Therefore, in principle, there is a point that the $\Lambda$ term dominates and so $\dot{G}=0$ and $\ddot{G}>0$, see also equation \eqref{FE3}. However one may naturally expect that $\Lambda$ becomes important only at the late times. Therefore it is very unlikely to have an anti-gravity to gravity transition at the early stages of the universe. Finally we deduce that $y$ and $r$ are positive quantities. More specifically, we shall show that if the evolution starts with suitable initial conditions including a positive $G$, $\dot{G}$ can be negative or positive during the cosmic evolution but $G$ remains positive.

Using the introduced dynamical variables, the equation of state parameter of dark energy, $\omega_\text{DE}$ and the effective equation of state parameter, $\omega_{\text{eff}}$, take the following simple form
\bea
\omega_{\text{DE}}&=&\!\! \frac{\!6 (5 \beta +14) r-24 m^2\!+\!132 x^2\!+\!78 y\!-\!(z\!-\!84) z\!-\!114}{90 \beta \! (r+y)\!-\!90} \nn\\
\omega_{\text{eff}}&=&\frac{1}{30} \left(-24 m^2+4 r-12 x^2+6 y-z^2+4 z+6\right)
\eea
Although $\omega_{\text{eff}}$ can be written just in terms of phase space coordinates, $\omega_{\text{DE}}$ contains a new variable $\beta = \frac{G_{\text{N}}}{G}$. It is clear form the definition of $\omega_{\text{DE}}$ that, in principle, the denominator can become zero. Consequently this parameter can become infinite. More specifically, this is the case for some initial conditions considered in this paper, for example see Fig \ref{newdeff}.

\subsection{MOG without cosmological constant($\Lambda$=0)}\label{sec-2}
As we have already mentioned, one of the main purposes of the current paper is to check if MOG can be considered as a dark energy model. To investigate if extra fields of MOG can play the role of dark energy, we set the cosmological constant to zero in the autonomous differential equations. It is equivalent to set $x=0$ in the equations of motion. In this case, the critical points $(y,r,m,z)$ of equations \eqref{yprime}-\eqref{zprime} are listed 
\bea\nn
&&\!\!\!\!\!p_{1,2}:(0,0,\pm \frac{\sqrt{24z - 24  - z^2}}{2 \sqrt6} ,z) ~~~~\omega_{\text{eff}}=1-\frac{2z}{3}\nn\\
&&\!\!\!\!\!p_{3}:(\frac{65}{27},0,0,-\frac{4}{3})\qquad \qquad ~~~~~~~~~~ \omega_{\text{eff}}=\frac{4}{9} \nn\\
&&\!\!\!\!\!p_{4}: ( \frac{391}{216},0,-\frac{\sqrt{3}}{4},-\frac{5}{6})\qquad  ~~~~~~~~~  \omega_{\text{eff}}=\frac{5}{18}\nn\\
&&\!\!\!\!\!p_{5}:(0, 1, 0,0) \qquad \qquad \qquad  ~~~~~~~~~ \omega_{\text{eff}}=\frac{1}{3} \nn\\
&&\!\!\!\!\!p_{6}: ( 0,\frac{81}{100},\frac{-1}{2\sqrt{3}},\frac{1}{5})\qquad  ~~~~~~~~~~~~  \omega_{\text{eff}}=\frac{4}{15}\nn\\
&&\!\!\!\!\!p_{7}: ( 0, 0,- \frac{5}{\sqrt{3}},2) \qquad \qquad  ~~~~~~~~~ \omega_{\text{eff}}=-\frac{19}{3} \nn\\
&&\!\!\!\!\!p_{8,9}: ( 0, 0,\pm \sqrt{\frac56},2) \qquad \qquad  ~~~~~~~ \omega_{\text{eff}}=-\frac{1}{3} \nn\\
&&\!\!\!\!\!p_{10,11} : ( 0, 0, \pm \sqrt{\frac{13}{8}}, 3)\qquad \qquad ~~~  \omega_{\text{eff}}=-1
\eea 
Surprisingly the fixed points are numbers and there is no free parameter to be constrained. This point is also clear from equations \eqref{yprime}-\eqref{mprime}, where with the aid of the special choice of the dynamical variables, free parameters do not appear in the autonomous differential equations. In fact, in principle, the free parameters appear in the coordinate of the fixed points \cite{Xu:2012jf}. Since each fixed point corresponds to an exact solution for the fields of the theory, existence of the free parameters in the fixed points provides a chance to make the model more consistent with the cosmological observations. However this is not the case in MOG.

In this sense MOG behaves like $\Lambda$CDM model where the fixed points are $(\Omega_R,\Omega_m,\Omega_{\Lambda})=$ $(1,0,0)$, $(0,1,0)$ and $(0,0,1)$, where $\Omega$'s are cosmic density parameters. Therefore, as we shall see MOG leads to clear cosmological consequences as in the case of $\Lambda$CDM model.
\begin{itemize}
	\item {\textbf{\textit{ $p_{1,2}$: $G$-$\mu$ dominated curves:}}}
\end{itemize}
 $p_1$ ($m > 0$) and $p_2$ ($m < 0$) correspond to two distinct
curves in the phase space. Every point on these curves
is a fixed point. In this case there is a constraint on
$z$ as $1.046 < z < 22.95$ to keep $m$ real. These curves
cover a wide range of $\omega_{\text{eff}}$ , from non phantom, to slightly
phantom and strongly phantom as $-14.3 \leq\omega_\text{eff}\leq 0.3$.
Eigenvalues of the stability matrix for $p_1$ and $p_2$ are
 $(0,2-z,3-z,\sqrt{2} -z\mp \sqrt{-(z-24) z-24})$, respectively.
It is easy to show that for $p_1$ one of the eigenvalues is positive for $z<3$ . Therefore, every point on the curve $p_1$ in this interval is unstable. On the other hand, in the case of $p_2$, for points in the interval $1.072<z<14.93$, there is at least one positive eigenvalue. Thus, these points are unstable. Note that, for $3 < z < 22.95$, sign of the eigenvalues of  $p_1$ are negative and reminders are zero. However, we can not simply decide that these points are stable. In fact, because of the existence of a zero eigenvalue, our first order perturbation analysis does not work and one has to use other methods, such as the center manifold theory, in order to reliably determine the stability of such a point. In the case of fixed points $p_{10}$ we have used the center manifold theorem and we write the results in the Appendix A.
\begin{itemize}
	\item {\textbf{\textit{ $p_3$: G-Matter dominated (GMD) era:}}}
\end{itemize}
This point corresponds to an expanding epoch in which the radiation density is zero and $\mu$ is constant. In other words, only matter and the scalar field $G$ dominate the evolution. Using the corresponding $\omega_{\text{eff}}$ one can easily show that $a(t) \propto t^{6/13}$, $G(t) \propto t^{-8/13}$ and $\phi_0 \propto t^{-18/13}$. In this phase the vector field mass $(\mu)$ is constant. It is noteworthy that the vector field's equation in MOG can be written as
\bea\label{vector}
\mu^2(t)=\frac{16 \pi \kappa}{\omega_0 \phi_0(t)}\rho_m(t)
\eea
Therefore the vector field mass directly depends on the energy density. This situation is reminiscent of chameleon scalar fields where the scalar field's mass depends on the environment's density. It is well known that this property leads to a screening effect for hiding the scalar fields effect in the local experiments \cite{Hinterbichler:2010es}. However, because of the appearance of $\phi_0(t)$ in the denominator of \eqref{vector}, it is not trivial to claim that such screening effect occur in MOG. Albeit screening effects are not just for scalar fields, and can happen in theories with vector fields \cite{BeltranJimenez:2013fca}. 
 It is clear that this solution $(p_3)$ is completely different from the standard matter dominated phase for which $a(t) \propto t^{2/3}$. Also, for $p_3$ eigenvalues are  $(-\frac{13}{6},-\frac{13}{6}, -1, -\frac{1}{3})$. Therefore, surprisingly $p_3$ is an attractive/stable critical point. A simple interpretation is that, in the context of MOG and in the absence of $\Lambda$, the universe can enter a permanent matter dominated era. This is grossly inconsistent with the cosmological observations that imply matter dominated epoch has been replaced by a stable dark energy dominated era \cite{Riess:1998cb}.
\begin{itemize}
	\item {\textbf{\textit{ $p_4$: $G \mu$-Matter dominated ($G \mu \text{MD}$) era:}}}
\end{itemize}
This critical point corresponds to an exact solution $a(t) \propto t^{12/23}$, $G(t) \propto t^{-10/23}$, $\mu(t) \propto t^{-18/23}$ and $\phi_0$ is constant. Interestingly this is exactly a solution that has been obtained in \cite{Roshan:2015uta} using the Noether symmetry approach. Also, $p_4$ is an unstable critical point since the eigenvalues are $(-2.28,-2.04,-1,0.26)$. From the stability point of view, it seems to be a true matter dominated era. We recall that, the radiation and matter eras are expected
to be unstable critical points. However $p_4$ is significantly different from the standard matter era for which the cosmic scale factor grows as $a(t) \propto t^{2/3}$. This situation is reminiscent of metric $f(R)$ cosmology where there is a "$\phi \text{MDE}$" regime for which the cosmic scale factor does not follow the standard behavior \cite{Amendola:2006we} and \cite{Amendola:2006kh}. We have a same fixed point ($f_3$ with a slightly different $\omega_{\text{eff}}$)
even in the presence of the $\Lambda$ term. We shall discuss more about these fixed points in the next section.
\begin{itemize}
	\item { \textbf{\textit{$p_5$: Radiation-dominated era:}}}
\end{itemize}
$p_5$ corresponds to a standard radiation dominated epoch
whose $\omega_{\text{eff}}=\frac{1}{3}$. Therefore the cosmic scale factor grows
as $a(t)\propto t^{1/2}$. In this era the vector field mass remains constant. Note that in $G\mu \text{MD}$ era this mass increases with time. Also the eigenvalues are $(-1, -1, 1, 0)$,
which establishes an unstable radiation era. Therefore, although there is no standard matter dominated phase in
MOG without $\Lambda$ ,there is a standard radiation era. It is also interesting that, unlike in the standard $\Lambda$CDM model, the expansion rate in the radiation dominated era is larger
than in the matter era ($p_3$).
\begin{itemize}
	\item {\textbf{\textit{ $p_6$: $G \mu$-Radiation dominated ($G \mu \text{RD}$) era:}}}
\end{itemize}
This point corresponds to an unstable $G \mu$-Radiation epoch for which the eigenvalues are $(-\frac{9}{5},1,-\frac{9}{10},\frac{9}{10})$. In this case, $a(t) \propto t^{10/19}$, $G(t) \propto t^{2/19}$ and $\mu(t) \propto t^{-10/19}$. Therefore unlike the matter dominated phases, $G(t)$ increases with time in this era. The behavior of this radiation dominated era is slightly different from the standard case. Note that for this fixed point we have $\omega_{\text{eff}}=\frac{4}{15}$.

\begin{figure}
\centerline{\includegraphics[width=8cm]{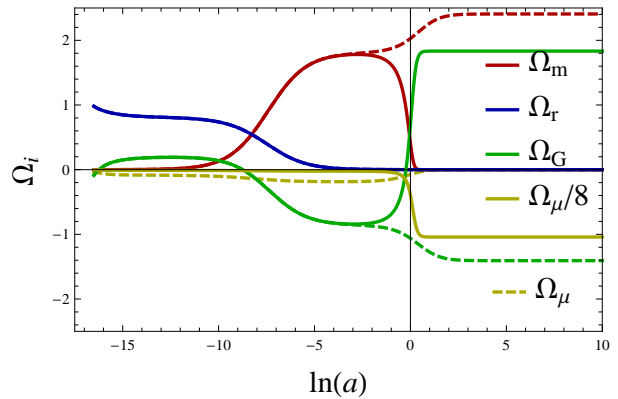}}
\caption[]{The cosmic evolution of $\Omega_i$ in the absence of $\Lambda$. The initial conditions are chosen deep in radiation dominated epoch, i.e. $z \approx 1.5 \times 10^7$ ($z$ is the redshift and should not be confused with the phase variable). The initial conditions for solid lines are $\Omega_R=0.98$, $\Omega_M =2.62716 \times 10^{-4} $, $\Omega_G \approx {-0.0863}$ and $\Omega_{\mu} =-0.0289 $. For dashed lines, we pick the same values except for  $\Omega_M$ which is $2.62717 \times 10^{-4} $. We see that for the later case system falls in the stable GMD point which is grossly inconsistent with the current cosmological observations.}
\label{Range}
\end{figure}

\begin{itemize}
	\item {\textbf{\textit{ $p_7$: Strongly Phantom attractor:}}}
\end{itemize}
Eigenvalues for this point are  $(-18,-17,-9,-9)$.
Therefore, $p_7$ shows a stable dark energy dominated era. $p_7$ corresponds to a strongly phantom behavior with $\omega_{\text{eff}} = \frac{-19}{3}$. It is easy to verify that the cosmic scale factor vary as $a(t)\propto (t_{\text{rip}}-t)^{-1/8}$. Where $t$ is smaller than the constant $t_{\text{rip}}$. In fact if $t=t_{\text{rip}}$, the Universe ends up with a finite-time. Although this is a stable dark energy dominated epoch, the strong phantom crossing is inconsistent with the observation, see \cite{Caldwell:2003vq} for more details. Also other fields grow as $G(t) \propto (t_{\text{rip}}-t)^{-1/4}$ and $\mu(t) \propto (t_{\text{rip}}-t)^{5/4}$. It is evident that phantom crossing can occur in MOG, see Fig. \ref{deefe}. It is important to mention that the universe may not enter this phase. In fact, the phase space trajectory of the system can end at $p_3$ before reaching the phantom attractor $p_6$. In Fig. \ref{Range} we have explicitly shown this fact. For some different initial conditions at the deep radiation dominated universe, we see that the dynamics reaches the stable $p_3$ point and stay there forever. On the other hand for a slightly different initial conditions the fixed point $p_6$ is realized. This fact explicitly shows that the dynamics starts from or close to an unstable point, i.e. $p_5$. Of course it is not needed to set the initial condition very close to that of solid lines in Fig. \ref{Range} in order to find $p_6$. For example if one set $\Omega_M$ to $3.7 \times 10^{-4} $, $p_6$ is still realized by the system. 

\begin{figure}
\centerline{\includegraphics[width=8.5cm]{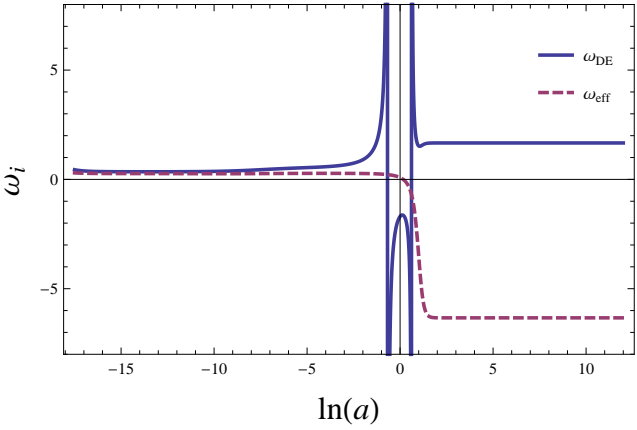}}
\caption[]{The evolution of $\omega_{\text{eff}}$ and $\omega_{\text{DE}}$ for the initial conditions presented in Fig. \ref{Range} for the solid curves. Although the effective equation of state parameter vary smoothly with time and shows the strongly phantom behavior, the dark energy's equation of state parameter experiences singularities at $r+y=\frac{1}{\beta}$.}
\label{newdeff}
\end{figure}

 More specifically, one may find initial conditions which lead to late time solutions $p_3$ or $p_7$. We have shown the system's evolution for such a set of initial conditions in Fig. \ref{Range}. Solid lines ends at strongly phantom attractor $p_7$. Note that the density parameters are related to our dynamical system variables as 
\bea
\Omega_m = y,\quad \Omega_R = r,\quad \Omega_{\Lambda} = x^2\quad\nn\\
     \Omega_G = z-\frac{z^2}{24},\quad \Omega_\mu = -m^2\quad 
\eea
Furthermore, in Fig. \ref{newdeff} we have shown the evolution of $\omega_{\text{eff}}$ and $\omega_{\text{DE}}$.

\begin{itemize}
	\item { \textit{\textbf{ $p_{8,9}$ : Unstable unaccelerated era:}}}
\end{itemize}
These critical points are on the lines $p_{1,2}$. In the case of $p_{8,9}$ the effective equation of state parameter is $-\frac{1}{3}$ and the eigenvalues are $(-2 \left(1 \mp \sqrt{10}\right),1,0,0)$ respectively. In this case, there is no acceleration and the cosmic scale factor grows uniformly with cosmic time. Also other functions vary as $G(t) \propto t^{2}$ and $\mu(t) \propto t^{\pm \sqrt{10}}$. It is worthy to mention that such a behavior for the cosmic scale factor is impossible in the context of $\Lambda$CDM model. In fact, in the standard model the acceleration $\ddot{a}(t)$ can vanishes only at a single moment. However, in MOG the effect of the extra fields can effectively appear as "repulsive" force which can eliminate the attractive nature of the gravity and provide a unaccelerated expansion. It is also somehow inconsistent with the expectation that MOG should lead to stronger force in he weak filed limit. We know that modified theories which try to address the flatness problem of the rotation curves of the spiral galaxies, have to strengthen the gravitational force. Albeit, it is necessary to stress that the evolution does not necessarily enter these epochs, i.e. $P_{8,9}$ for a cosmologically viable trajectory.
\begin{itemize}
	\item { \textbf{\textit{$p_{10,11}$: Unstable de Sitter-like era:}}}
\end{itemize}
These points are also on the lines $p_{1,2}$. The effective equation of state parameter for these points is $\omega_{\text{eff}} = -1$. Therefore $p_{10,11}$ correspond to an epoch in which the cosmic scale factor grows exponentially. The scalar fields vary with time as $G(t)\propto e^{3 t} $ and $\mu(t)\propto e^{\pm t\sqrt{39/2}} $. Also the eigenvalues are $(-3\mp \sqrt{78},-1,0,0)$ . One may certainly conclude that $p_{11}$ is an unstable fixed point. On the other hand, the stability of non-hyperbolic point $p_{10}$ can be shown using center manifold theory. In the Appendix A we use this theory to specify the stability character of $p_{10}$. Also in Fig. \ref{poincaree} we showed that $p_{10}$ is also an unstable critical point. Therefore, there is no late time stable de Sitter phase in the cosmic evolution of MOG, when $\Lambda$ is zero.

Now let us summarize the general cosmological behavior of MOG in the absence of the cosmological constant (or equivalently when the scalar field $G$ is not massive). In this case, it seems that MOG does not possesses a true consequences of cosmological phases. The fixed points are: unstable radiation dominated $(p_5)$ or unstable $G \mu \text{RD}$ $(p_6)$, unstable $G\mu\text{MD}$ point $(p_4)$ followed by the late time strongly phantom attractor $p_6$. There is an interesting feature in the dynamics of this model. In fact there is a matter dominated attractor $G\text{MD}$, i.e. point $p_3$. In other words, regarding the initial conditions, the universe can enter this matter dominated phase and stay there forever. Although, one can choose initial conditions for which the evolution does not realize $p_3$, the late time attractor $p_7$ also is not physically accepted. In Fig. \ref{Range}, we have plotted the evolution of the density parameters for two set of initial condition. In the absence of $\Lambda$, MOG can not be considered  as a dark energy model in the sense that its extra fields can not play the role of dark energy. Also, as we have already mentioned, the standard matter dominated era is replaced with the $G\mu\text{MD}$ epoch at which the scale factor grows as $a(t)\propto t^{12/23}$. Clearly this behavior is far away from the standard case. We will discuss more about this important point in the next section.
\subsection{MOG with cosmological constant ($\Lambda \neq 0$)}\label{sec-3}
In this section we explore the original version of MOG, i.e. $\Lambda\neq 0$. In this case, $x'\neq 0$ and we use the same dynamical variables introduced in section \ref{sec-2} and find the relevant fixed points $(y,r,m,z,x)$. Setting to zero the right hand side of equations \eqref{yprime}-\eqref{mprime} we find the following critical points: 
\bea
&&\!\!\!\!\!f_{1,2}  :  ( 0, 0,\pm \frac{ \sqrt{-24 +\! 24 z\! -\! z^2}}{2 \sqrt{6}},z, 0) ~~\omega_{\text{eff}}=1 - \frac{2 z}{3}\nn\\
&&\!\!\!\!\!f_{3} :(\frac{65}{27},0,0,-\frac{4}{3},0)
\qquad \qquad ~~~~~~~ \quad \omega_{\text{eff}} = \frac{4}{9}\nn\\
&&\!\!\!\!\!f_{4} : (\frac{391}{216},0,-\frac{\sqrt{3}}{4},-\frac{5}{6},0) ~~~~~~~~~~~~~~~ \omega_{\text{eff}}=\frac{5}{18}\nn\\
&&\!\!\!\!\!f_{5} : ( 0,  1, 0,0,0) ~~~~~~~~~~~~ \quad \qquad \qquad \omega_{\text{eff}}=\frac{1}{3}\nn\\
&&\!\!\!\!\!f_{6} : (0,\frac{81}{100},\frac{-1}{2\sqrt{3}},\frac{1}{5},0) ~~~~~~~~~~~~~~~~~~ \omega_{\text{eff}}=\frac{4}{15}\nn\\
&&\!\!\!\!\!f_{7} : (0, 0, 2,-\frac{5} {\sqrt{3}}, 0)\qquad \qquad ~~~~~~~~~\omega_{\text{eff}}=-\frac{19}{3}\nn\\
&&\!\!\!\!\!f_{8,9} : ( 0, 0, \pm \sqrt{\frac{5}{6}}, 2, 0) \quad ~~~~~~~~~~~~~\quad  \omega_{\text{eff}}=-\frac13\nn \\
&&\!\!\!\!\!f_{10,11} : ( 0,  0 , \pm \sqrt {\frac{13}{8}} , 3, 0)~~~~~~~~~ \qquad \omega_{\text{eff}}=-1\nn\\
&&\!\!\!\!\!f_{12}: (0,0,0,-\frac{4}{3},\pm \sqrt{\frac{65}{27}}) \qquad \quad ~~~~~~~ \omega_{\text{eff}}=-1\nn\\
&&\!\!\!\!\!f_{13,14}\!\! : (0,0, \frac{-11}{7 \sqrt{3}}, \frac{-2}{7}, \pm \frac{2}{7} \sqrt{\frac{46}{3}})~~~~~~~ \omega_{\text{eff}}=-1\nn
\eea
In what follows we shall study the stability of the above mentioned fixed points and discuss their physical interpretation. We emphasize again that, we are checking the possibility that if MOG can be a cosmologically viable theory.

$f_{1,2}$ indicate two different curves in the phase space. Every point on these curves is a fixed point and the relevant eigenvalues are the same as lines $p_{1,2}$ in the previous section. On the other hand, $f_3$ is similar to the $G\text{MD}$ phase (where the matter and the scalar field $G$ dominate the evolution i.e. $p_3$) introduced in the previous section. However, it is interesting that unlike the $p_3$ point, $f_3$ is an unstable fixed point. In other words, existence of the cosmological constant change the character of this fixed point. Eigenvalues of the stability matrix are $({-\frac{13}{6},-\frac{13}{6},\frac{13}{6},-1,-\frac{1}{3}})$ and the cosmic scale factor grows as $a(t)\propto t^{6/13}$ and for the other fields we have $G(t) \propto t^{-8/13}$, $\mu$ is constant and $\phi_0 \propto t^{-18/13}$. We confront again the question that does this point correspond to a suitable matter dominated era? If not, then this fact put a serious doubt on the cosmological  validity of this model even if it pass the local experiments and address the dark matter problem in galactic scale.

\begin{itemize}
	\item {\textbf{\textit{$f_{4}$: $G\mu\text{MD}$ era:}}}
\end{itemize}
This point is similar to $G\mu\text{MD}$ point $p_4$. The eigenvalues are $(-3.37,-\frac{23}{12},\frac{23}{12},1.45,-1)$. Therefore $f_4$ is an unstable critical point. Furthermore, the scale factor grows as $a(t)\propto t^{12/23}$ and for other functions we have $\mu(t) \propto t^{-18/23} $, $ \phi_0 $ is constant and $G(t) \propto t^{-10/23}$. It is interesting that there is two possible matter dominated phases in MOG, i.e. $f_3$ and $f_4$. The expansion rate in $f_4$ is slightly faster than $f_3$ but still too slower than the standard matter dominated case. We emphasize that this point is the main result of the current paper. The behavior of MOG in the matter dominated era is crucial because this theory is an alternative theory for dark matter particles. Therefore, it should possesses an appropriate matter era in which structure formation happens without any need to cold dark matter particles. 

As we mentioned before, this fact may put a serious doubt on the viability of this model. In fact a slower expansion rate, in principle, changes the duration of the matter dominated phase. Consequently, there may be some impacts on the cosmic microwave background (CMB) observations, for example on the angular size of the  sound horizon. On the other hand, the strength of the gravitational force is different in MOG than the standard case. Therefore the growth rate of matter perturbations will be different from that of $\Lambda$CDM model. So observational data of galaxy clustering may help to distinguish the consequences of MOG. To summarize, existence of a non-standard expansion rate in the matter dominated epoch put a serious constraint on the viability of MOG but does not necessarily rule out MOG. One need to explore the cosmic structure formation in the context of MOG and carefully check the impacts of this theory on the CMB observations, in order to make a reliable decision about the viability of the theory \cite{us}.

\begin{itemize}
	\item {\textbf{\textit{$f_{5}$: Radiation-dominated era:}}}
\end{itemize}
This point is an standard unstable radiation dominated era for which $\omega_{\text{eff}}=\frac{1}{3}$. Also the relevant eigenvalues are: $(2,-1,-1,1,0)$. We recall that even in the absence of $\Lambda$ there was an unstable radiation dominated era $p_5$. These points, namely $p_5$ and $f_5$, are the only eras at which the scalar fields are constant. It is worthy to mention that at these epochs the vector field varies rapidly with time, as $\phi_0 \propto t^{-2}$, since it is directly coupled to the radiation density. However after this phase the "running" of the scalar fields starts and eventually they dominate the evolution. Note that $f_6$ also have the same behavior as $p_6$, i.e. an unstable $G \mu \text{RD}$ era which is slightly different from the standard radiation era since $\omega_{\text{eff}}=\frac{4}{15}$. One can choose initial condition in a way that starting point of the evolution of the universe be either $f_5$ or $f_6$.
\begin{figure}
\centerline{\includegraphics[width=8.5cm]{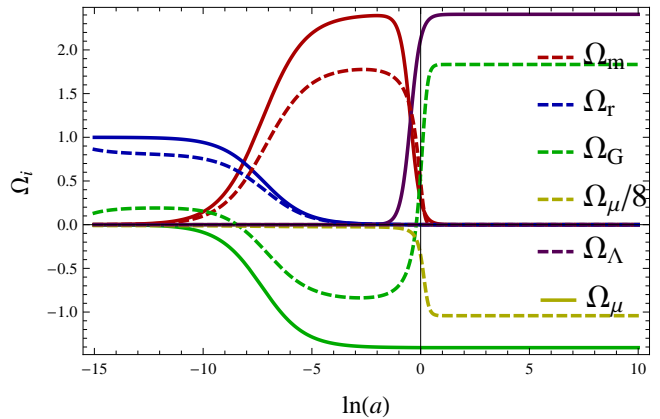}}
\caption[]{The evolution of $\Omega_i$ for two set of initial conditions. The initial conditions for the solid lines are:  $\Omega_{M} =10^{-4} $, $\Omega_R = 0.999$, $\Omega_x = 10^{-30}$, $\Omega_G  \approx 0.99 \times 10^{-4}$ and $\Omega_{\mu} =-10^{-8}$ at $z = 3.43 \times 10^7$. One can see a true sequence of the cosmological epochs for this choice of initial conditions. The initial conditions for dashed lines are $\Omega_M =2.62716 \times 10^{-4} $, $\Omega_R = 0.98$, $\Omega_x = 10^{-30}$, $\Omega_G \approx -0.086 $ and $\Omega_\mu =-0.0289$ at $z \approx 10^7$. Solid lines reaches the de-sitter like attractor $f_{12}$ through the $f_3$ while dashed lines ends at strongly phantom attractor $f_{7}$ passing $f_4$. A common feature in the dynamics of these different initial conditions is that the scalar fields remain constant during the early stages of the universe.}
\label{omegatwo}
\end{figure}
\begin{itemize}
	\item {\textbf{\textit{$f_{7}$: Strongly Phantom attractor:}}}
\end{itemize}
In this case eigenvalues are $(-18, -17, -9, -9, -8)$. Therefore, $f_{7}$ is a stable critical point. This point is exactly similar to $p_7$ presented in the previous section. Thus all our analysis for $p_7$ are also true for $f_{7}$. We just mention that, $f_{7}$ is different from a standard phantom crossing era and it can not be considered as an acceptable late time solution, because of the strongly phantom crossing behavior. The initial conditions will specify the final fate of the system, i.e. $f_{12}$ or $f_{7}$. It is interesting that MOG provides two kinds of late time accelerated expansions, de Sitter or a strongly phantom solution. 

\begin{figure}
\centerline{\includegraphics[width=8cm]{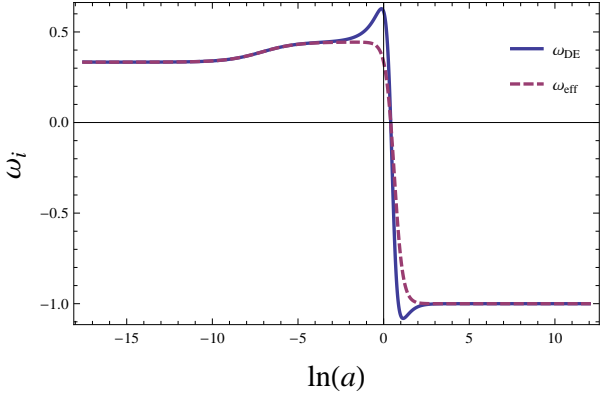}}
\caption[]{The evolution of $\omega_{\text{DE}}$ and $\omega_{\text{eff}}$. The initial conditions are the same as for Fig. \ref{omegatwo} for solid lines. The phantom crossing behavior is clear.}
\label{deefe}
\end{figure}

\begin{figure}
\centerline{\includegraphics[width=8.5cm]{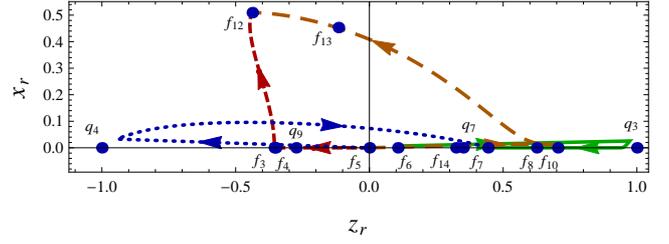}}
\caption[]{The phase space of the system in poincar\'{e} coordinate.  The dashed (red) line shows a cosmological path that starts from radiation dominated epoch  $f_{5}$, passing trough matter dominated epoch, $f_{3}$ and ends at stable de sitter-like attractor $f_{12}$. The dotted blue line shows a cosmological evolution starting from radiation dominated $f_5$ passing through  $q_4$ and reaches at $f_{7}$. The long-dashed line shows that universe starting from radiation dominated  $f_6$ and passing through unstable points $f_{7,14}$ and $f_{8},f_{10}$, toward unstable era $f_{13}$ and ends at a late time attractor $f_{12}$. The green solid line also starts from $f_6$, passing through $q_7$ and $q_3$, ends at strongly phantom $f_{7}$. Note that there are unstable fixed points at infinity, $q_{3,4}$ and $q_{7,9}$.}
\label{poincaree}
\end{figure}

Furthermore, $f_{8,9}$ are the same as the unaccelerated solutions $p_{8,9}$. In this case eigenvalues are $(-2 \mp 2 \sqrt{10},1,1,0,0)$ respectively, which show that $f_{8,9}$ are unstable points. For these solutions the scale factor uniformly increases with time. Our analysis in the previous section for $p_{8,9}$ is also true for these solutions. Also $f_{10,11}$ with eigenvalues $(-3\mp \sqrt{78},-1,0,0,0)$ are similar to $p_{10,11}$. It is clear that $f_{11}$ is unstable and $f_{10}$ is a non hyperbolic fixed point. We tried to check the stability of $f_{10}$ using the center manifold theory as we did for $p_{10}$. However, unlike $p_{10}$, we realized that even center manifold theorem does not reveal the character of these point. Therefore one needs to use more advanced methods such as \textit{the normal form theory}. On the other hand we are considering a five dimensional manifold and these methods become very complicated and are out of the scope of this paper. Thus we have to rely to our phase space trajectories to decide about the stability of this point. We see in Fig. \ref{poincaree} that $f_{10}$ appears as an unstable critical point. This is also consistent with the result that we have already obtained for $p_{10}$. Note that $f_{8,9}$ and $f_{10,11}$ are special cases of the lines $f_{1,2}$ with $z=2$ and $z=3$, respectively.

\begin{itemize}
	\item {\textbf{\textit{$f_{12}$: de Sitter-like attractor:}}}
\end{itemize}
This point corresponds to the stable dark energy dominated universe where the dynamics is dominated with the cosmological constant and the scalar field $G$. Eigenvalues are $(-\frac{16}{3},-\frac{14}{3},-\frac{13}{3},-\frac{13}{3},-\frac{13}{3})$ and the cosmic scale factor grows exponentially. Also this point corresponds to the exact solution $G(t) \propto e^{-4t/3}$ and $\mu$=constant. We recall that in the absence of the cosmological constant there is no stable dark energy dominated era with $\omega_{\text{eff}}=-1$.

\begin{itemize}
\item {\textbf{\textit{$f_{13,14}$: $G\mu\Lambda$ era:}}}
\end{itemize}
These points correspond to an epoch at which the scalar fields together with the cosmological constant dominates the evolution and the ordinary matter and radiation energy densities are zero. The cosmic scale factor grows exponentially and for other functions we have $G(t) \propto e^{-2t/7}$ and $\mu(t) \propto e^{-22 t/7} $ . Also the relevant eigenvalues are $(-6.03,-\frac{30}{7},-\frac{23}{7},-\frac{23}{7},2.74)$ which clearly show that these points are unstable. Obviously these points can not be considered as early time unstable fixed points (i.e. points which can show an inflationary period). Because the contribution of the cosmological constant is comparable to the other components while we know that $\Lambda$ does not play an important role in the early universe. Also one may expect that for a very early time unstable fixed point all the eigenvalues should be positive. It is also clear in Fig. \ref{poincaree} that $f_{14}$ is neither a late time fixed point nor an early time one.
  
A true cosmological path could start from an unstable radiation dominated epoch, $f_5$ or $f_6$ (which differs slightly from the standard radiation dominated epoch), continue toward an unstable matter dominated epoch, $f_3$ or $f_4$ and finish in a stable dark energy dominated point, $f_{12}$. Note that  strongly phantom attractor, $f_{7}$ is not acceptable as a standard late time solution. It is clear form Fig. \ref{poincaree} that for specific initial conditions phase trajectory of the system passes through unaccelerated eras $f_{8}$ and $f_{10}$. However this trajectory is not acceptable because it does not realize the matter dominated points. Also we have plotted the behavior of $\omega_{\text{DE}}$ and $\omega_{\text{eff}}$ in Fig. \ref{deefe}. It is important to mention that although $\omega_{\text{eff}}$ remains always greater than $-1$, $\omega_{\text{DE}}$ becomes smaller than $-1$ about present time. In other words, MOG can provide a slightly phantom solutions at present time.

Let us summarize the main results of this section. We explicitly showed that MOG possesses true cosmological sequence of the cosmological epochs. In fact there is a standard radiation dominated point $f_5$. Also there are two unstable matter dominated phases which are not standard in the sense that the cosmic scale factor grows slowly than the standard case. On the other hand there is a stable late time solutions $f_{12}$.

\section{Phase space analysis at infinity}
\label{sec-4}
It is important to stress that, form equation \eqref{FE1} one may straightforwardly conclude that $H$, in principle, can become zero. In such a moment, our dynamical variables become infinite. Therefore it seems necessary to check this possibility. In other words, we have to check the fixed points lying in the infinity of the dynamical system. As we shall show, such a study will ensure us that there is nothing special, such as a regular \textit{bounce}, at the moment when the expansion rate of the universe become zero. We know that even if we find some fixed points in the infinity, it does not mean that a true cosmological trajectory will realize that points. However, finding these points and analyzing their stability will provide a better understanding for the general behavior of the system. This section is devoted to explore this issue. We show that there are  unstable fixed points in the infinity of this model.

We know that the phase space defined with the dynamical variables \eqref{var} is not closed. In other words, our variables can take infinite values. In order to check the behavior of the system at infinity, we introduce new Poincar\'{e} coordinates $x_r, y_r, r_r, z_r$ and $ m_r $ as follow
\bea\label{poincar}
\begin{split}
&x=\frac{x_r}{\sqrt{1-R_r^2}},\,\,\,\,y=\frac{y_r}{\sqrt{1-R_r^2}},\,\,\,\,z=\frac{z_r}{\sqrt{1-R_r^2}}\\&
r=\frac{r_r}{\sqrt{1-R_r^2}},\,\,\,\,m=\frac{m_r}{\sqrt{1-R_r^2}}
\end{split}
\eea
where $R_r=\sqrt{m_r^2+r_r^2+x_r^2+y_r^2+z_r^2}$. Therefore for points at infinity we have $R_r=1$ . In the following, after rewriting the autonomous equations \eqref{yprime}-\eqref{mprime} in the new coordinates and defining a new "time" $\lambda$ as $dx/d\lambda=(1-R_r^2)x'$, we have taken the limit $R_r \rightarrow 1$. Finally the autonomous equations at infinity take the following forms
\bea\label{n1}
\begin{split}
&\frac{dy_r}{d\lambda}=\frac{1}{20} y_r (40 \sqrt{3} m_r (y_r+z_r+r_r) (x_r^2+y_r^2+z_r^2+r_r^2-1)\nn\\&
+m_r^2 (4 (12 x_r^2+9 y_r^2+4 r_r^2-9)-24 (3 y_r+2 r_r) z_r
-71 z_r^2)\nn\\&
+\!\!40 \sqrt{3} m_r^3 (y_r+z_r+r_r)\!-\!z_r^2 (23 x_r^2\!+\!58 y_r^2+78 r_r^2\!-\!58)\!\nn\\&
-\!24 (3 y_r+2 r_r) z_r (x_r^2+y_r^2+r_r^2-1)+4 (r_r^2 (13 x_r^2+y_r^2-1)\nn\\
&+3 (6 x_r^2 (y_r^2\!-\!1)\!+\!4 x_r^4\!+\!y_r^4\!-\!2 y_r^2\!+\!1)-2 r_r^4)\nn\\&
-24 (3 y_r+2 r_r) z_r^3-71 z_r^4)\nn\\&
\frac{1}{z_r}\frac{dz_r}{d \lambda}=\frac{1}{y_r}\frac{dy_r}{d \lambda} \!+\!\frac{6 (3 y_r\!+\!2 r_r) (m_r^2\!+\!x_r^2\!+\!y_r^2\!+\!r_r^2\!-\!1)}{5 z_r}
\nn\\&
+\frac{3}{5} (9 m_r^2+8 x_r^2+7 y_r^2\!+\!7 r_r^2\!-\!7)\!+\!\frac{6}{5} (3 y_r\!+\!2 r_r) z_r\!+\!\frac{1}{4} 17 z_r^2\nn\\&
\frac{1}{m_r}\frac{dm_r}{d \lambda}=\frac{1}{y_r}\frac{dy_r}{d \lambda}+\frac{m_r (48 x_r^2\!+\!36 y_r^2\!+\!37 z_r^2\!+\!36 r_r^2\!-\!36)}{20 m_r} 
\nn\\&
-\frac{(y_r\!+\!z_r\!+\!r_r) (-\!60 m_r^3\!+\!40 \sqrt{3}( m_r^2\!+\!x_r^2\!+\!y_r^2\!+\!z_r^2\!+\!r_r^2\!-\!1))}{20 m_r}\nn\\&
\frac{1}{r_r}\frac{dr_r}{d \lambda}= \frac{1}{y_r}\frac{dy_r}{d \lambda} +m_r^2+x_r^2+y_r^2+z_r^2+r_r^2-1\nn\\&
\frac{1}{x_r}\frac{dx_r}{d \lambda}= \frac{1}{y_r}\frac{dy_r}{d \lambda} - \frac{3}{5} \left(x_r^2+2 y_r^2+2 r_r^2-2\right)-\frac{23 z_r^2}{20}
\end{split}
\eea
Setting to zero the right hand side of these equations and keeping in mind that fixed point at infinity are constrained as $m_r^2+r_r^2+x_r^2+y_r^2+z_r^2=1$, we find the following critical points $( y_r,r_r, m_r, z_r,x_r )$
\bea
\begin{split}
&~~q_1=(1,0,0,0,0) \\& ~~q_2=(0,1,0,0,0)\\&
q_{3,4}=(0,0,0,\pm1,0)\\& q_{5,6}=(0,0,0,0,\pm1)\\
& q_{(7-10)}=(0,0,\pm 5 \sqrt{\frac{2}{57}},\pm \sqrt{\frac{7}{57}},0)
\end{split}
\label{fp}
\eea
Note that concerning the constraint \eqref{non}, the only accepted fixed points are $q_{3,4}$ for which $z_r=-1$ and $q_{7-10}$. In fact one may rewrite the constraint \eqref{non} as 
\begin{equation}
\begin{split}
(y_r + r_r +z_r)&\sqrt{1+R^2}\\& + (x_r ^2 -m_r ^2 -z_r^2/24)(1+R^2) \leq 1 
\end{split}\label{non2}
\end{equation}
where $R=\sqrt{m^2+r^2+x^2+y^2+z^2}$. In order to check the existence of points $q_1$-$q_{10}$ we divide the inequality \eqref{non2} by $(1+R^2)$, and take the limit of $R \rightarrow \infty$. In this case we have 
\begin{equation}
(x_r^2 -m_r^2 -\frac{z_r^2}{24}) \leq 0
\end{equation}
Also if $x_r=m_r=z_r=0$ (namely for $q_1$ and $q_2$), one may divide \eqref{non2} by $\sqrt{1+R^2}$ and take the limit of $R \rightarrow \infty$. In this case we find
\begin{equation}
(y_r +r_r) \leq 0
\end{equation}
Therefore our fixed points should satisfy the relevant constraint. Now it is straightforward to verify that the only allowed fixed points at infinity are $q_{3,4}$ and $q_{7-10}$. The corresponding eigenvalues of the stability matrix constructed from equations \eqref{n1} for $q_{3,4}$ and $q_{7-10}$ are $(\frac{1}{10},-\frac{1}{20},-\frac{1}{20},0,0)$ and $(\frac{1207}{570},-\frac{1207}{1140},-\frac{1207}{1140},0,0)$, respectively. Note that due to symmetry of equations, the eigenvalues of stability matrix for $q_{7}$ to $q_{10}$ are the same. This means that accepted fixed points are unstable critical point. 

One may provide a simple interpretation for these point. In fact they are "middle" time, and not an early or late time, point where the scalar fields $G$ and $\mu$ dominates the evolution and the expansion rate becomes zero for a moment. In other words if $q_{3,4}$ or $q_{7-10}$ was a stable point, then the universe could enter an static phase and stay there forever. In Fig \ref{poincaree} we have shown two different trajectories which realize this point. More specifically the blue dotted curve starts from the radiation dominated point $f_5$ and after passing the infinity fixed points $q_9$ and $q_4$, falls into the late time point $f_{7}$. The green solid line shows another path which starts from $G \mu \text{RD}$  point, $f_6$, passes the unstable infinity fixed points $q_7$ and $q_3$, and eventually reaches the strongly phantom stable point $f_{7}$.

\section{conclusion}
\label{sec-4.1}
In this paper we have considered the cosmological behavior and consequences of a scalar-vector-tensor theory of gravity, known as MOG in the literature. Although this theory is known as an alternative theory for dark matter particles, we have investigated it's viability as a dark energy model using the so-called dynamical system method. In fact, we checked the possibility that if the extra fields of MOG can play the role of dark energy. We first derived the autonomous equations of the relevant dynamical system, i.e. equations \eqref{yprime}-\eqref{mprime}, and found the corresponding fixed points in two different cases: $\Lambda=0$ and $\Lambda\neq 0$. 

In section \ref{sec-2}, we showed that in the absence of $\Lambda$, there is not a standard late time epoch. More specifically, the evolution starts from the unstable radiation dominated epoch $p_5$, then reaches the unstable matter dominated epoch $p_4$ and eventually ends at the stable late time strongly phantom attractor $p_7$, which is not physically acceptable. It is worth mentioning that depending on initial conditions, universe could enter the stable $G\text{MD}$ era ($p_3$). Obviously such an initial condition leads to a wrong cosmological behavior. Although $p_5$ is a standard radiation dominated era, the matter dominated epoch $p_4$ and the late time attractor $p_7$ are not standard. In fact in $p_4$ the scale factor grows as $a(t)\propto t^{\frac{12}{23}}$ that is slower than the standard case in which $ a(t)\propto t^{\frac{2}{3}}$. On the other hand, for the late time attractor $p_7$, $\omega_{\text{eff}}$ is $-\frac{19}{3}$, which is not a standard late time solution. It shows that in the absence of $\Lambda$, one can not recover standard cosmological epochs in MOG.

In section \ref{sec-3}, we investigate the original form of the theory, i.e. with nonzero $\Lambda$. It is noteworthy that in this theory the cosmological constant $\Lambda$ can be considered as the mass of scalar field $G$ (or equivalently $\chi$), see equation \eqref{action}. Also note that there are two coupling constants $\omega_0$ and $\kappa$ in this theory. Surprisingly these parameters do not appear in our phase space analysis. In other words, our fixed points are numbers and there is no free parameter to be constrained. In this sense MOG behaves like $\Lambda$CDM model where the fixed points do not include any free parameter. 

The first effect of the nonzero cosmological constant is to change the character of the  stable matter dominated phase $p_3$ (or equivalently $f_3$). In other words, two matter dominated epochs of MOG, namely $f_3$ and $f_4$ are now unstable as expected. Interestingly, there are also two early time radiation dominated and $f_5$ and $f_6$ and also two late time attractors $f_4$ and $f_{7}$. In fact, depending on the initial conditions, universe could start from standard radiation era $f_5$ or $G \mu \text{RD}$ epoch, $f_6$, which is slightly different from the standard radiation dominated era, continues toward matter dominated epochs $f_3$ or $f_4$ and end at strongly phantom, non physical, attractor $f_{7}$ or the standard stable de-sitter epoch $f_4$. Also, the phantom crossing behavior in MOG is clearly seen, as shown in Fig. \ref{deefe}. As we already mentioned, a true cosmological path could start from standard radiation dominated epoch $f_5$, continue toward one of the unstable matter dominated $G\text{MD}$ ($f_3$) or $G\mu\text{MD}$ ($f_4$), and finally reach the stable late time solution $f_4$ as shown in Fig \ref{omegatwo}. 

Furthermore, there is a new feature in the cosmology of MOG that is absent in the standard model. In fact there are unstable eras $p_{8,9}$ (or equivalently $f_{8,9}$) in which the universe expands uniformly. In these eras the existence of the extra fields, in large scales, behave like an effective repulsive "force" and cancel out the attractive nature of the gravitation, and consequently cosmic acceleration vanishes. 

As a final remark, we emphasize that there is no standard matter dominated phase in the model of MOG studied in this paper. In fact with or without cosmological constant the scale factor grows as $a(t)\sim t^{0.5}$ instead of the conventional $t^{2/3}$. In principle, this may lead to inconsistencies with CMB and large scale structure formation observations. Therefore, it seems necessary to investigate this issue with more careful considerations. We leave this issue as a future study. It is important to mention that even if the above mentioned observations rule out the original form of MOG, one may find some special models of MOG (with different self interaction potentials) which can pass the observations.

\section{Acknowledgments}
This work is supported by Ferdowsi University of Mashhad under Grant No. 39640(04/11/1394). We thank John Moffat, Viktor Toth, Martin Green and especially Fatimah Shojai for insightful comments. Sara Jamali thanks Luca Amendola and Sohrab Rahvar for valuable discussions.

\appendix
\section{stability of $p_{10}$ using Center Manifold Theory}
\label{sec-5}
\begin{figure}
\centerline{\includegraphics[width=6.5cm]{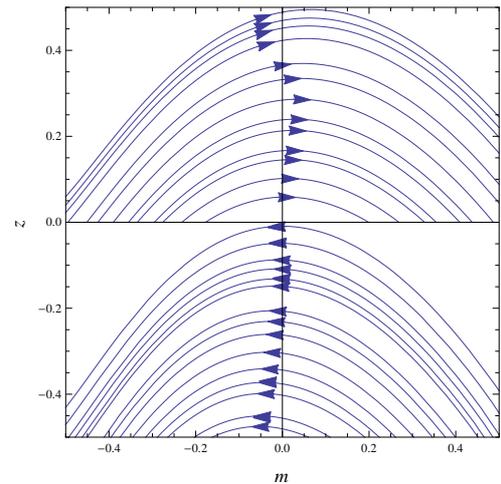}}
\caption[]{The phase space plot of center manifold of point $p_{10}$. Instability of the center is clearly seen. }
\label{cm}
\end{figure}
The stability of a \textit{hyperbolic} point $\mathbf{x_0}$ of the non-linear system $\mathbf{\dot{x}=f(x)}$ can be determined by the behavior of the linear system $\mathbf{\dot{x}=A x}$, where $\mathbf{A=Df(x_0)}$ and $\mathbf{D}$ denotes differentiation with respect to $\mathbf{x}$. On the other hand, the stability of a \textit{non-hyperbolic} point, can be determined by recognizing the behavior of the \textit{center manifold} near that point \cite{perko}. In other words, since zero eigenvalues reveals no information about the qualitative and stability behavior of the system, one should find a way to get information about this part of the system.

 Considering the phase space of MOG, we find the non-hyperbolic fixed point $p_{9}$. For the sake of simplicity, we change the coordinate in a way that $p_9$ lies at the origin. More specifically, the simple change of coordinates as  $z \rightarrow z+3 $ and $m \rightarrow m+\sqrt{\frac{13}{8}}$ would transfer $p_9$ to the origin. In this case the transformed autonomous equations are
\bea
y^{\prime}&=&-\frac{1}{5} 12 m^2 y-\frac{6}{5} \sqrt{26} m y+\frac{2 r y}{5}+\frac{3 y^2}{5}-\frac{y z^2}{10}+\frac{4 y z}{5}\nn\\
r^{\prime}&=&\!-\!\frac{1}{5} 12 m^2 r\!-\!\frac{6}{5} \sqrt{26} m r\!+\!\frac{2 r^2}{5}\!+\!\frac{3 r y}{5}\!-\!\frac{r z^2}{10}\!+\!\frac{4 r z}{5}\!-\!r\nn\\
z^{\prime}&=&-\frac{6 m^2 z}{5}-\frac{6 m^2}{5}-\frac{3}{5} \sqrt{26} m z-\frac{3 \sqrt{26} m}{5}+\frac{r z}{5}\nn\\
&-&\frac{9 r}{5}+\frac{3 y z}{10}-\frac{27 y}{10}-\frac{z^3}{20}+\frac{17 z^2}{20}+\frac{9 z}{10}\\
m^{\prime}&=&\!-\!\frac{6 m^3}{5}\!-\!2 \sqrt{3} m^2\!-\!\frac{9}{5} \sqrt{\frac{13}{2}} m^2\!+\!\frac{m r}{5}\!+\!\frac{3 m y}{10}\!-\!\frac{m z^2}{20}\nn\\
&+&\frac{9 m z}{10}\!-\!\sqrt{78} m\!-\!\frac{39 m}{10}\!+\!2 \sqrt{3} r\!+\!\frac{1}{10} \sqrt{\frac{13}{2}} r\!+\!2 \sqrt{3} y\nn\\
&+&\frac{3}{20} \sqrt{\frac{13}{2}} y \!-\!\frac{1}{40} \sqrt{\frac{13}{2}} z^2\!-\!\frac{z^2}{4 \sqrt{3}}\!+\!\frac{3 \sqrt{3} z}{2}\!+\!\frac{9}{20} \sqrt{\frac{13}{2}} z\nn
\eea

In order to apply center manifold theory, one also should build a box diagonal matrix of linear coefficient of the equations. This means that the system of equations  should be separated in two distinct parts; the equations which have  negative eigenvalues and those with zero eigenvalues in the stability matrix. Performing this separation for complicated systems, like the equations we are considering, one have to apply the Jordan transformation to the system of equations. For this case, it can be applied using the following transformation:
\bea
&z& \!\! \rightarrow \frac{80}{27}\! \sqrt{\frac{2}{13}} m\!+\! \sqrt{\frac{72}{13}} r\!+\!\frac{2}{161} \!\! \sqrt{6\! \left(\!239\!-\!20 \sqrt{78}\right)} y\!\!+\!\!\frac{ \sqrt{104} z}{3}\nn\\
&r& \rightarrow  2 \sqrt{\frac{2}{13}} r,\,\,m  \rightarrow   r\!+\!y\!+\!z,\,\,\, y   \rightarrow  \frac{1}{9} (-20) \sqrt{\frac{2}{13}} m
\eea
The resulting equations in complete form are too long to be written here, although the calculations are straightforward. Nevertheless, let us write some terms
\bea\label{cmprime}
y^{\prime} &=& (-3-\sqrt{78}) y + \mathcal{O}(y^2+ yr+..)\nn\\
r^{\prime} &=&  -r + \mathcal{O}(r^2 + ry+ ..)\\
z^{\prime}&=& m +  \mathcal{O}(z^2+zm+..)\nn\\
m^{\prime}&=& -\frac{\sqrt{26}}{3} m z -10 \sqrt{\frac{2}{13}} m r  + ...\nn
\eea
 Now, the general form of the equations, for zero and negative eigenvalues, are in matrix form
 \bea
\it{X}^{\prime}= C \it{X} + F(\it{X},\it{Y})\nn\\
\it{Y}^{\prime}= P \it{Y} + K(\it{X},\it{Y})\nn
\eea
$C$, in our case,  is a $2 \times 2$ matrix which contains the linear coefficients of equations  with  zero eigenvalues and $P$ has the same role for equations with negative real parts of eigenvalues. $F $ and $K$ are $2 \times 1$  matrices that denote the rest terms of equations which does not included in $C$ and $P$, respectively. In fact, using the Jordan decomposition, one can separate the linear and nonlinear parts of equations. Here, investigating the linear part of transformed equations, one can easily find out that $y$ and $r$ have negative eigenvalues, while $z$ and $m$ have zero eigenvalues in the stability matrix. In order to clarify the behavior of zero components of the system using the center manifold theory, we expand the negative components with respect to zero ones as $y = a z^2 + b m z + c m^2 + ...$ and $r = d z^2 + e m z + f m^2+...$, see \cite{perko} for more details. Following the relevant theorems, one finds that there exists a $2$ dimensional invariant center manifold for which:
\[ 
   \begin{pmatrix} y \\ r \end{pmatrix}=
  \begin{pmatrix} h_1 \\ h_2 \end{pmatrix}
\]
where functions $h_{1,2}=h_{1,2}(z,m)$ satisfy the following equality
$$ Dh [C (^{\,z}_{m})  + F]- P h - K =0 $$
or equivalently in the the matrix form
 \[ 
   D\begin{pmatrix} h_1 \\ h_2 \end{pmatrix}\!
   \left[\!
   \!\begin{pmatrix} c_1 & c_2 \\ c_3 & c_4 \end{pmatrix} 
  \!\begin{pmatrix} z \\ m \end{pmatrix}+
  \!\begin{pmatrix} f_1 \\ f_2 \end{pmatrix} \right]\!=\!
  \!\begin{pmatrix} p_1 & p_2 \\ p_3 & p_4 \end{pmatrix}
  \!\begin{pmatrix} h_1 \\ h_2 \end{pmatrix}+
  \!\begin{pmatrix} k_1 \\ k_2 \end{pmatrix}
\]
Note that in this matrix equation, $D$ denotes differentiating with respect to $z$ and $m$ (components with zero eigenvalues in the Jordan form) and so $Dh(z,m)$ is a $2 \times 2$ matrix. Noting equation \eqref{cmprime}, one finds that
 \[
\!\begin{pmatrix} c_1 & c_2 \\ c_3 & c_4 \end{pmatrix}=
\!\begin{pmatrix} 0 & 1 \\ 0 & 0 \end{pmatrix} \qquad
\begin{pmatrix} p_1 & p_2 \\ p_3 & p_4 \end{pmatrix}=
\begin{pmatrix} -3-\sqrt{78} & 0 \\ 0 & -1 \end{pmatrix}
\] 
 Where $F = F(z,m,h(z,m))$ and $K = K(z,m, h(z,m))$.  Solving this matrix equation, one finds the expansion coefficient, $ a, b, ..$ and can find out the behavior of center manifold using the following equation
$$\it{X}^{\prime}= C \it{X} + F(\it{X} ,h(\it{X})) $$
As already mentioned, $\it{X}$ stands for the zero components. It is noteworthy that in the expansion of $y$ and $r$ with respect to $z$ and $m$, we look for the first nonzero coefficient, which will determine the behavior of the center manifold. Performing this method for $p_{10}$, one finds  
\begin{equation}
\begin{split}
& m^{\prime} \approx 1.63228 m^2 z^2+1.17611 m^2 z\\& ~~~~~~+0.406759 m^2-5.08273 m z^2-3.39935 m z\\& z^{\prime} \approx -0.342063 m^2 z^2-0.357055 m^2 z\\& ~~~~~~+0.138952 m^2+1.49451 m z^2+2.03504 m z+m\nn
\end{split}
\end{equation}
Using these equation, in Fig. \ref{cm}, we have plotted a two dimensional phase space in the $z-m$ plane. One may straightforwardly conclude that the origin, $p_{10}$, represents an unstable critical point.


\begin{thebibliography}{99}


\bibitem{moffat2}
 J.~W.~Moffat,
  JCAP {\bf 0603}, 004 (2006)

\bibitem{Moffat:2013uaa} 
  J.~W.~Moffat and S.~Rahvar,
  Mon.\ Not.\ Roy.\ Astron.\ Soc.\  {\bf 441}, no. 4, 3724 (2014)
  
    
\bibitem{Brownstein:2005zz} 
  J.~R.~Brownstein and J.~W.~Moffat,
  Astrophys.\ J.\  {\bf 636}, 721 (2006)  
  
  \bibitem{Brownstein:2005dr} 
  J.~R.~Brownstein and J.~W.~Moffat,
  Mon.\ Not.\ Roy.\ Astron.\ Soc.\  {\bf 367}, 527 (2006)
  
  
  
  \bibitem{Milgrom:1983ca} 
  M.~Milgrom,
  Astrophys.\ J.\  {\bf 270}, 365 (1983).
  
      \bibitem{Bekenstein:2004ne} 
  J.~D.~Bekenstein,
  Phys.\ Rev.\ D {\bf 70}, 083509 (2004).
  
    \bibitem{Moffat:2014pia}
  J.~W.~Moffat and V.~T.~Toth,
  Phys.\ Rev.\ D {\bf 91} (2015) 4,  043004.
  
  
  
  
  \bibitem{Roshan:2015gra}
  M.~Roshan and S.~Abbassi,
  Astrophys.\ J.\  {\bf 802} (2015).

  \bibitem{ra}
  M.~Roshan and S.~Abbassi,
  Phys.\ Rev.\ D {\bf 90}, no. 4, 044010 (2014)
   
  \bibitem{Roshan:2015uta}
  M.~Roshan,
  Eur.\ Phys.\ J.\ C {\bf 75} (2015) .
  
  \bibitem{Moffat:2014bfa} 
  J.~W.~Moffat,arXiv:1409.0853
  
  \bibitem{Moffat:2015bda} 
  J.~W.~Moffat,
  arXiv:1510.07037; J.~R.~Mureika, J.~W.~Moffat and M.~Faizal,
  arXiv:1504.08226; J.~W.~Moffat,
  Eur.\ Phys.\ J.\ C {\bf 75}, no. 3, 130 (2015).
  
  
  
 \bibitem{Xu:2012jf} 
  E.~J.~Copeland, A.~R.~Liddle and D.~Wands,
  Phys.\ Rev.\ D {\bf 57}, 4686 (1998).
   L.~Amendola,
  Phys.\ Rev.\ D {\bf 62}, 043511 (2000).
  C.~Xu, E.~N.~Saridakis and G.~Leon,
  JCAP {\bf 1207}, 005 (2012). S.~Capozziello and M.~Roshan,
  Phys.\ Lett.\ B {\bf 726}, 471 (2013).
   J. Wainwright,  G.F.R. Ellis, Dynamical Systems in Cosmology,(Cambridge University Press, 1997); O.~Hrycyna and M.~Szydlowski,
  JCAP {\bf 1012}, 016 (2010); M.~Rinaldi,
  JCAP {\bf 1510}, no. 10, 023 (2015);
 M.~Roshan and F.~Shojai,
  Phys.\ Rev.\ D {\bf 94}, 044002 (2016)
  
  
 \bibitem{Roshan:2014mqa}
  M.~Roshan,
  Phys.\ Rev.\ D {\bf 87}, no. 4, 044005 (2013)
 \bibitem{Alnes:2005rw} 
  H.~Alnes, M.~Amarzguioui and O.~Gron,
  Phys.\ Rev.\ D {\bf 73}, 083519 (2006)
 
 
 \bibitem{GarciaBellido:2008yq} 
  J.~Garcia-Bellido and T.~Haugboelle,
  JCAP {\bf 0909}, 028 (2009).
    
  \bibitem{perko}
  L. Perko, \textit{Differential equations and dynamical systems},( Springer,  1991)
  
  \bibitem{Faraoni}
 V. Faraoni, \textit{Cosmology in scalar tensor gravity}, (Springer, 2004)
 
 \bibitem{anti}
 I. Bars and A. James, Phys. Rev. D{\bf 93}, no. 4, 044029
(2016); 
   K. Bamba, Shin'ichi Nojiri, S. D. Odintsov, D. Saez-Gomez , Phys. Lett. B {\bf730}, 136 (2014);  J. J. M. Carrasco,
W. Chemissany and R. Kallosh, JHEP {\bf 1401}, 130 (2014).
   
 
 \bibitem{Hinterbichler:2010es} 
 J. Khoury and A. Weltman, Phys. Rev. Lett. 93, 171104
(2004); J. Khoury and A. Weltman, Phys. Rev. D {\bf69},
044026 (2004).
  
  
  \bibitem{BeltranJimenez:2013fca} 
  J.~B.~Jimenez, A.~L.~Delvas Froes and D.~F.~Mota,
  Phys.\ Lett.\ B {\bf 725}, 212 (2013)
 
 \bibitem{Riess:1998cb} 
 A.~G.~Riess {\it et al.} [Supernova Search Team Collaboration],
  Astron.\ J.\  {\bf 116}, 1009 (1998)
  
  \bibitem{Amendola:2006we}
  L.~Amendola, R.~Gannouji, D.~Polarski and S.~Tsujikawa,
  Phys.\ Rev.\ D {\bf 75} (2007) 083504
  
  \bibitem{Amendola:2006kh} 
  L.~Amendola, D.~Polarski and S.~Tsujikawa,
  Phys.\ Rev.\ Lett.\  {\bf 98}, 131302 (2007)
  
  \bibitem{Caldwell:2003vq} 
  R.~R.~Caldwell, M.~Kamionkowski and N.~N.~Weinberg,
  Phys.\ Rev.\ Lett.\  {\bf 91}, 071301 (2003)
  
  \bibitem{us}
 Sara Jamali and Mahmood Roshan, work in progress 
\end{thebibliography}
\end{document}